\documentclass[sn-mathphys,Numbered]{sn-jnl}

\usepackage{graphicx}
\usepackage{dcolumn}
\usepackage{bm}
\usepackage{dcolumn}
\newcolumntype{d}[1]{D{.}{.}{#1}}

\usepackage{multirow}%
\usepackage{amsmath,amssymb,amsfonts}%
\usepackage{amsthm}%
\usepackage{mathrsfs}%
\usepackage[title]{appendix}%
\usepackage{xcolor}%
\usepackage{textcomp}%
\usepackage{manyfoot}%
\usepackage{booktabs}%
\usepackage{algorithm}%
\usepackage{algorithmicx}%
\usepackage{algpseudocode}%
\usepackage{listings}%

\renewcommand{\vec}[1]{\boldsymbol{#1}} 

\usepackage{color}
\definecolor{Blue}{rgb}{0.3,0.3,0.9}
\definecolor{Red}{rgb}{0.9,0.3,0.3}
\definecolor{Green}{rgb}{0.3,0.6,0.3}

\newcommand{\revision}[1]{#1}

\newif\ifNOSUP \NOSUPfalse

\begin{document}


\title{
%
Quantum engineering for compactly localized states in disordered Lieb lattices
}

\author*[1]{\fnm{Carlo} \sur{Danieli}}
\email{carlo.danieli87@gmail.com}
\affil*[1]{\orgdiv{Institute for Complex Systems}, \orgname{National Research Council} (ISC-CNR), \orgaddress{\street{Via dei Taurini 19}, \postcode{00185} \city{Rome}, \country{Italy}}}

\author[2,3]{\fnm{Jie} \sur{Liu}}
\email{liujie@csuft.edu.cn}
\affil[2]{\orgdiv{School of Physics and Optoelectronics}, \orgname{Xiangtan University}, \orgaddress{\city{Xiangtan} \postcode{411105}, \country{China}}}
\affil[3]{\orgdiv{Institute of Mathematics and Physics}, \orgname{Central South University of Forestry and Technology}, \orgaddress{\city{Changsha}  \postcode{410004}, \country{China}}}


\author[2,4]{\fnm{Rudolf A.} \sur{R\"omer}}
\email{r.roemer@warwick.ac.uk}
\affil[4]{\orgdiv{Department of Physics}, \orgname{University of Warwick}, \orgaddress{\street{Gibbet Hill Road}, \city{Coventry}, \postcode{CV4 7AL}, \country{United Kingdom}}}

\date{\today}
\abstract{
Blending ordering within an uncorrelated disorder potential in families of 3D Lieb lattices preserves the macroscopic degeneracy of compact localized states and yields unconventional combinations of localized and delocalized phases -- as shown in \href{https://journals.aps.org/prb/abstract/10.1103/PhysRevB.106.214204}{Phys.Rev.B {\bf 106}, 214204 (2022)}. 
We proceed to reintroduce translation invariance in the system by further ordering the disorder, and discuss the spectral structure and eigenstates features of the resulting perturbed lattices. 
We restore order in steps by first (i) rendering the disorder binary -- {\it i.e.} yielding a randomized checkerboard potential, then (ii) reordering the randomized checkerboard into an ordered one, and at last (iii) realigning all the checkerboard values yielding a constant potential shift, but only on a sub-lattice. 
Along this path, we test the influence of additional random impurities on the order restoration. 
We find that in each of these steps, sub-families of states are projected upon the location of the degenerate compact states, while the complementary ones are localized in the perturbed sites with energy determined by the strength of checkerboard.
This strategy, herewith implemented in the 3D Lieb, highlights order restoration as experimental pathway to engineer spectral and states features in disordered lattice structures in the pursuit of quantum storage and memory applications. 
}



    

    


\maketitle

\section{Introduction}

Quantum memories, as emerging quantum technology applications, hold immense potential, but they are still confronted with significant challenges \cite{Lvovsky2009OpticalMemory,Julsgaard2004ExperimentalLight}.  
Recently, some pioneering work has proposed a possible promising implementation of quantum network storage based on so-called compact localized (eigen)states (CLS) \cite{Rontgen2019}.
These CLS occupy each only a \revision{finite and typically small physical subset of the lattice network} for an extended duration, endowing them with excellent storage properties \cite{Rontgen2019}. 
\revision{Furthermore, CLS in translationally invariant networks are macroscopically degenerate and form a {\it dispersionless} (or {\it flat}) band in the Bloch structure. Hence, lattices supporting CLS go by the name of {\it flat band} networks~\cite{Derzhko2015a,Leykam2018,Leykam2018c}.
The existence of CLS relies on fine-tuning the hopping terms of a lattice~\cite{Dias2015a,Maimaiti2017a,Rontgen2018a}.  
Likewise, achieving the storage and transfer of CLS relies on the control via fine-tuned driving of coupling terms in a \revision{flat band} lattice \cite{Rontgen2019}.}   

The CLS are unaffected by changes to the system outside their compact localization domain. However, when such changes are introduced within the domain, \revision{typically} they quickly loose their localization confinement. These changes are indeed easily possible via environmental  influences that lead to, e.g., fluctuations in the values of the local terms of the system Hamiltonian. Hence, local disorder can quickly destroy the CLS \cite{Leykam2013,Flach2014a,Leykam2017,Mao2020b,Liu2020a}. On the other hand, the established knowledge of the spatial structure of the CLS in a given lattice structure also allows to devise special disorder potentials that help to confine the CLS to their original domain \cite{Bodyfelt2014,Danieli2015,Nandy2015b,Xia2018,Liu2022}. The idea is to engineer the potentials such that the non-CLS domains become energetically less favourable, hence suppressing the delocalization and reestablishing the confinement in the CLS domain. 

In the present work, we shall study the CLS as present in a 3D Lieb lattice, $\mathcal{L}_3(1)$, which is a 3D generalization of the original Lieb lattice \cite{Lieb1989a,Mielke1991a,Tasaki1992a,Mielke1993}, and part of a much larger family of such lattices \cite{Zhang2017b,Mao2020b,Liu2021}. Instead of directly manipulating the CLS, we target the non-CLS dispersive states, transforming them gradually with external potentials to exhibit similar energy and spatial behavior as the CLS. We choose the potential fluctuations such that they are negligible on the sites available for the CLS. But on the non-CLS sites, we impose fluctuations with ever more decreasing levels of disorder, i.e., (i) by allowing only two potential values, (ii) by forcing those two values to follow a spatially alternating, i.e.\ 3D-checkerboard-like arrangement and (iii), by performing a simple constant potential shift. One might also usefully think of this sequence as an introduction of more and more ``order'' in the potential fluctuations.
While there are of course quantitative differences in the spectral properties between these three cases, we find that, even in the third, most simple case, about $50\%$ of the dispersive states become more CLS-like: they localize predominantly on the domains of the CLS, enhancing the viability of using CLS for long-term quantum memory storage. Naturally, the CLS remain unaffected throughout.
We then also allow additional random disorder fluctuations on top of the underlying three orders. When added as a perturbation, this additional disorder does not alter the outcome of our study: the non-CLS states that become more CLS-like still do so and hence are still enhancing the stability of the CLS.

Section \ref{sec:lieb-lattices} provides an overview of the recent results for the $\mathcal{L}_d(n)$ Lieb lattices and explains the structure of the potentials used to perform the eventual non-CLS confinement. The numerical measures of spatial confinement and localization are explained in section \ref{sec:method} while results for the three cases of potentials are given in sections \ref{sec:random-checkerboard}, \ref{sec:ordered-checkerboard} and \ref{sec:constant-shift}. We conclude in section \ref{sec:conclusions}.

\section{Lieb lattices and ordering restoration}
\label{sec:lieb-lattices}

The Lieb lattices $\mathcal{L}_d(n)$ are a parametric family of systems which extend the standard $d$-dimensional orthogonal lattices by including $n$ equispaced atoms between each two neighboring vertices. 
As discussed in Ref.~\cite{Liu2022}, these systems possess $d n + 1$ Bloch bands, among which $(d-1) n$ are strictly flat bands and the corresponding eigenstates are spatially compact, {\it i.e.}, have strictly non-zero amplitude within a finite volume of the system. Paradigmatic cases of this family are the $\mathcal{L}_2(1)$ Lieb lattice -- a 2D system obtained with a single additional atom between each two originally neighboring vertices, which has been intensely studied theoretically \cite{Zhang2017b,Leykam2018,Leykam2018c,Mao2020b} and has been  experimentally realized with ultracold atoms~\cite{Shen2010,Goldman2011,Apaja2010} and photonic waveguide networks~\cite{Mukherjee2015a,Vicencio2015a,Guzman-Silva2014} -- as well as its 3D version, $\mathcal{L}_3(1)$ \cite{Mao2020b}. 
These two notable lattices, and those obtained for $n=2,3$ and $4$, have been studied in the presence of uncorrelated disorder potential in Refs.~\cite{Mao2020b,Liu2020a,Liu2021}. 
The latter $d=3$ ones have then been further investigated in Ref.~\cite{Liu2022} in the presence of a mix of order and disorder which preserves the macroscopically degenerate compact eigenstates -- resulting in a projection of the majority of states upon the sub-space spanned by the CLSs, while keeping the remaining states Anderson localized and spread across the reminder of the energy spectrum. 
%
\begin{figure}[tb]
    \centering
    (a)\includegraphics[width=0.4\textwidth]{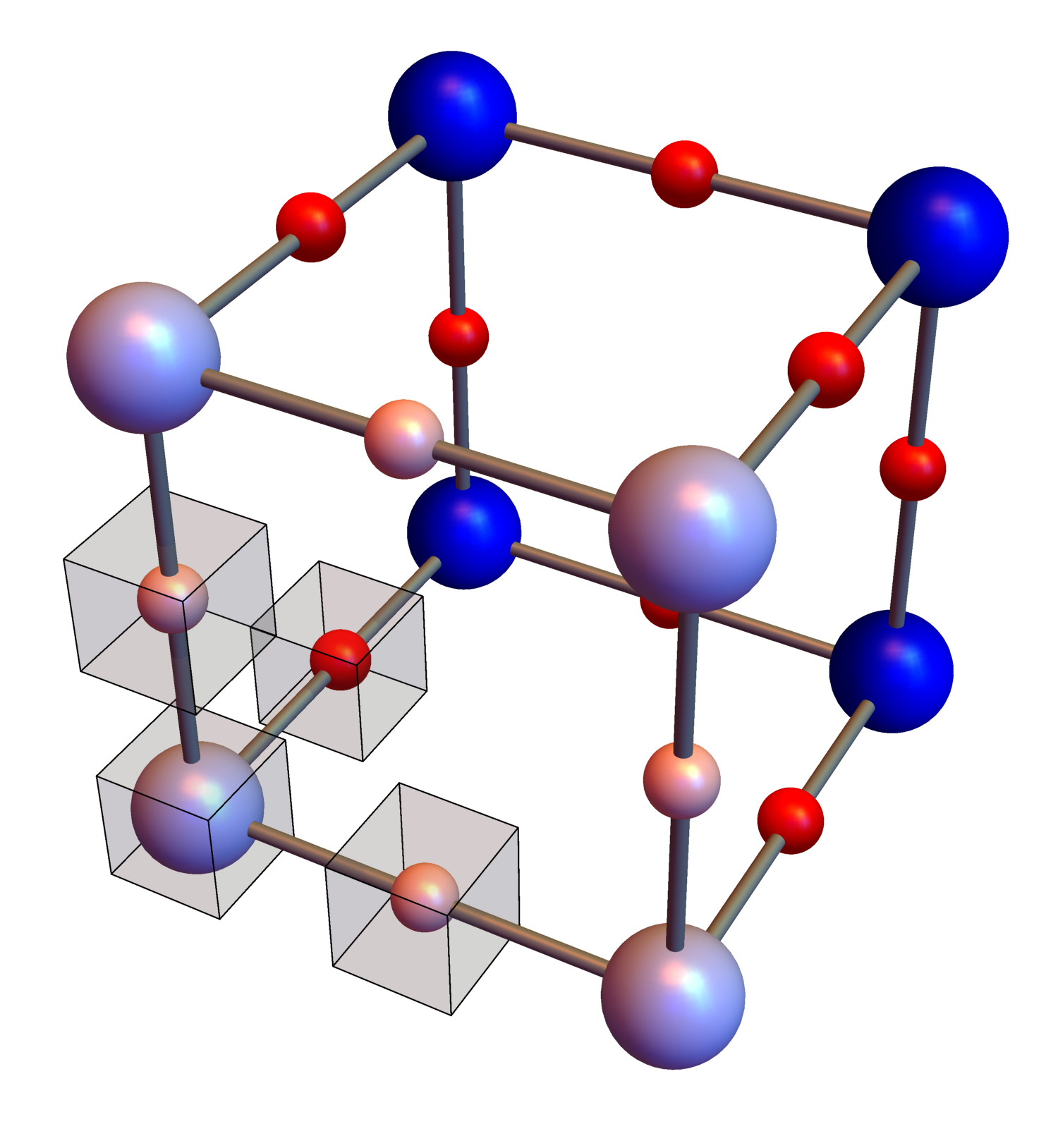}
    (b)\includegraphics[width=0.45\textwidth]{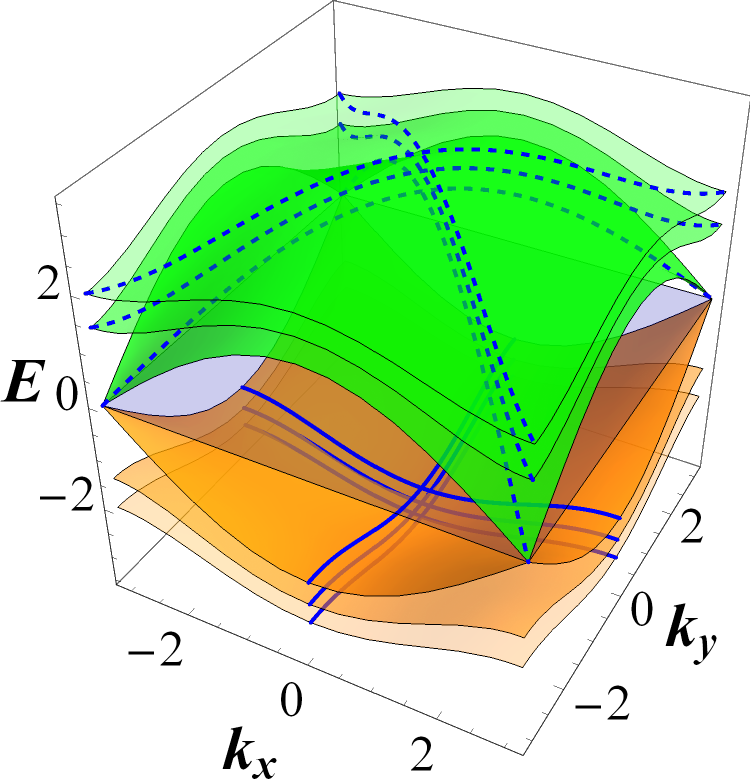}
    \caption{(a) A section of the Lieb lattice $\mathcal{L}_3(1)$ 
    distinguishing between cube ($\mathcal{C}$) (blues) and Lieb ($\mathcal{L}$) sites (reds). The semitransparent gray cubes indicate the minimal unit cell, while the black lines represent the hopping connections. The light blue and red spheres towards the reader indicate a face of $\mathcal{L}_3(1)$ supporting compactly localised states.
    (b) Dispersive energy bands $E_{D+}$, $E_{D-}$ for the clean system, with  $E_{D+}$ in green colours and $E_{D-}$ in orange ones. Increasing levels of opaqueness indicate $k_z=0$, $\pi/2$ and $\pi$ with the latter showing a Dirac-like cone at $(k_x, k_y, k_z)=(\pm \pi, \pm \pi, \pi)$ and $(\pm \pi, \mp \pi, \pi)$.
    The dashed lines on $E_{D+}$ highlight $k_x=\pm k_y$ while the solid lines for $E_{D-}$ denote $k_x=0$ or $k_y=0$.
    The doubly degenerate flat band at $E=0$ is shown in blue.
    }
    \label{fig:Lieb_schematic}
\end{figure}

We now further explore the impact of mixing order and disorder in the localization features of the 3D Lieb models. 
Therefore, we focus solely on the prototypical $\mathcal{L}_3(1)$ Lieb lattice -- schematically represented in Fig.~\ref{fig:Lieb_schematic}(a) -- and consider a mixing between order and random potential which progressivelys restore order in the system. 
The lattice Hamiltonian reads
    \begin{equation}
        H=\sum_{\vec{r}} \varepsilon(\vec{r}) |\vec{r}\rangle \langle \vec{r}|
        - \sum_{\vec{r}\neq \vec{r'}}t_{\vec{r}\vec{r'}}|\vec{r}\rangle \langle \vec{r'}|\ ,
    \label{Equ:def1} 
    \end{equation} 
where $|\vec{r}\rangle$ indicates the orthonormal Wannier state for an electron situated at site $\vec{r}=(x,y,z)$. 
The set of real numbers $\varepsilon_{\vec{r}}$ represents the onsite potential. 
We set the hopping integrals $t_{\vec{r}\vec{r'}} \equiv 1$ for nearest-neighbor sites $\vec{r}$ and $\vec{r'}$, and $t_{\vec{r}\vec{r'}} \equiv 0$ otherwise, following the lattice profile shown in  Fig.~\ref{fig:Lieb_schematic}(a). 
The four bottom left corner sites of the lattice, enclosed in cubes, form the minimal unit cell of the lattice. 
Hence, in absence of onsite energies, i.e.\ $\varepsilon(\vec{r}) = 0\  \forall\ \vec{r}$, the \emph{clean} Lieb lattice $\mathcal{L}_3(1)$ posses four energy bands 
\begin{equation}
    E_{F1,F2} =  0, \quad
    E_{D\pm}  =  \pm\sqrt{6+2(\cos k_x + \cos k_y + \cos k_z)},
\label{eq:bands}
\end{equation} 
where $k_x,k_y,k_z$ are the reciprocal vectors in the momentum space. 
The two bands $E_{F1}$, $E_{F2}$ are macroscopically degenerate at $E=0$, giving effectively a single doubly degenerate flat band. 
Each of the CLS corresponding to the flat bands is enclosed within a 2D square plaquette of the network (cp.\ Fig.~\ref{fig:Lieb_schematic}(a)), and has equal amplitudes and opposite phases arranged clockwise along the four added sites \cite{Zhang2017b,Liu2022}.
The two dispersive bands $E_{D\pm}$ are clearly symmetric with respect to $E=0$, and in Fig.~\ref{fig:Lieb_schematic}(b) we show 2D surfaces of the 3D Bloch band structure from Eq.~\eqref{eq:bands} for fixed $k_z=0,\pi/2$ and $\pi$. In particular, we observe that both dispersive bands are touching the flat bands $E=0$ at the point $R = (\pi, \pi, \pi)$ of the Brillouin zone. As discussed in Ref.~\cite{Liu2020a}, around the touching point $R$, the dispersive bands form a linear cone, with a depleting density of states at $E=0$. 
\begin{figure}
    \centering
    \includegraphics[width=0.95\textwidth]{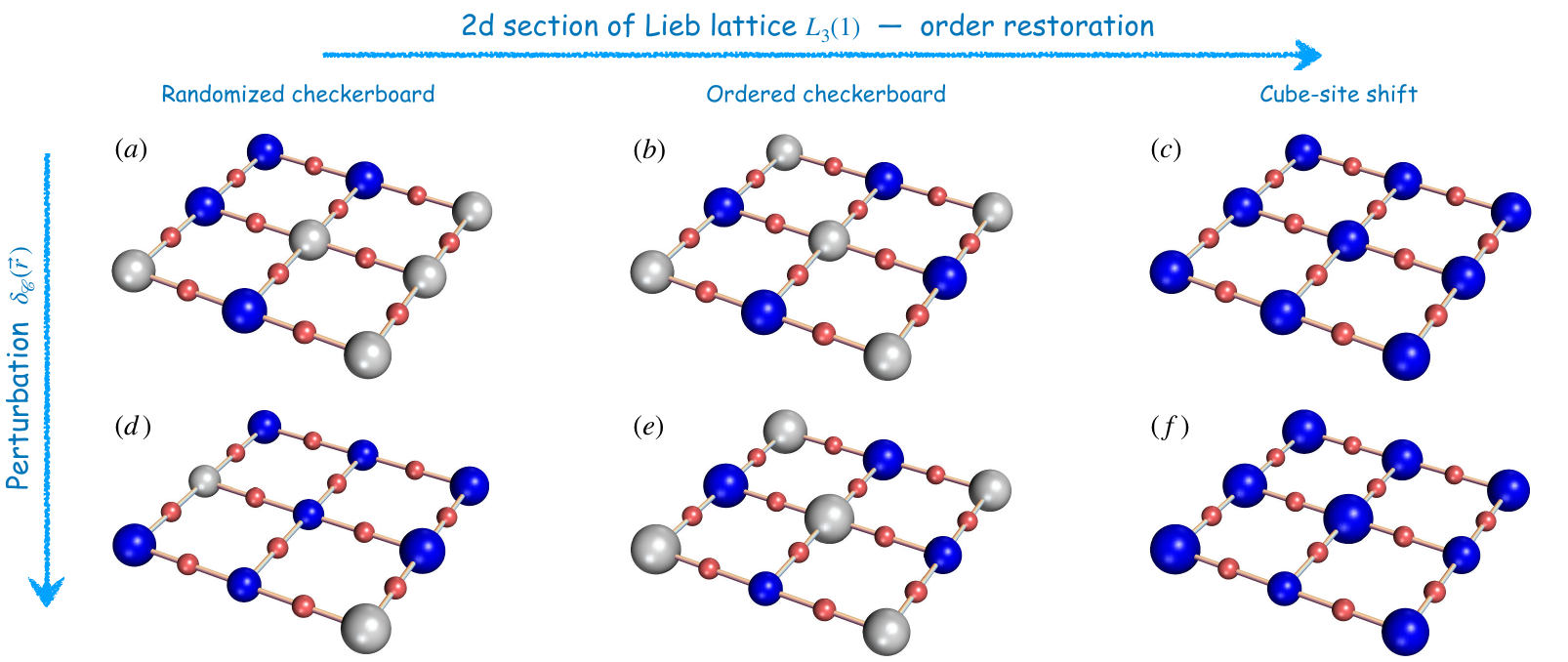}
    \caption{Schematic representation of the order restoration in the Lieb model $\mathcal{L}_3(1)$, shown for convenience in a 2D section of the 3D geometry. As in Fig.~\ref{fig:Lieb_schematic}, the sites are distinguished between Lieb sites (red) and cube sites, which herewith are split in blue for $\varepsilon_\mathcal{C}(\vec{r}) = +\mu$ and gray for $\varepsilon_\mathcal{C}(\vec{r}) = -\mu$. The size of the blue/gray spheres indicate the magnitude of the potential ({\it i.e.} the strength $\mu$ in absolute value). 
    (a) randomized checkerboard arrangement of the cube potential $\varepsilon_\mathcal{C}(\vec{r}) = \pm \mu$, (b) ordered checkerboard arrangement of the cube potential $\varepsilon_\mathcal{C}(\vec{r}) = \pm \mu$, and (c) constant potential shift on cube sites $\varepsilon_\mathcal{C}(\vec{r}) = + \mu$. 
    (d-f) Same as (a-c), with additional perturbation $\varepsilon_\mathcal{C}(\vec{r}c) \longmapsto \varepsilon_\mathcal{C}(\vec{r}) + \delta_\mathcal{C}(\vec{r})$ as indicated by the slightly irregular size of the blue/gray spheres). 
    }
    \label{fig:dir_to_ord}
\end{figure}

As a bipartite network~\cite{Ramachandran2017}, we split the Lieb lattice \eqref{Equ:def1} in two sub-lattices: the {\it cube} sub-lattice $\mathcal{C}$ formed by the sites located at the vertexes colored in blue in Fig.~\ref{fig:Lieb_schematic}(a), and the {\it Lieb} sub-lattice $\mathcal{L}$ formed by those sites, colored in red, sitting in between two blue vertexes in Fig.~\ref{fig:Lieb_schematic}(a).
In Ref.~\cite{Liu2022}, we then considered the simplest choice of \emph{correlated potentials} $ \varepsilon_{\vec{r}}$ which preserve the macroscopically degenerate CLS -- {\it i.e.} setting the Lieb site potential $\varepsilon_\mathcal{L}(\vec{r})$ constant, \emph{e.g.} without loss of generality,
$\varepsilon_{\mathcal{L}}(\vec{r})\equiv 0$, while introducing a non-zero uncorrelated random potential $\varepsilon_{\mathcal{C}}(\vec{r})$ on the cube sites. 
This choice of onsite potential projects most of the eigenstates onto the macroscopically degenerate compact states at $E = 0$. Such projection induces a separation in the energy spectrum between states, with energies $E\approx 0$ and delocalized amplitudes within the Lieb sub-lattice $\mathcal{L}$, and a remainder of the eigenstates that have high energies and are Anderson localized within the cube sub-lattice $\mathcal{C}$. 

We now further delve inside this mix of order and disorder, aiming especially to outline the simplest conditions that yield the projection of the eigenstates onto the preserved CLSs.
We do this alongside the previously considered potential in Ref.~\cite{Liu2022}, namely zeroing the onsite potential $\varepsilon_{\mathcal{L}}(\vec{r})\equiv 0$ of Lieb sites, and iteratively restore the order in the uncorrelated randomized potential $\varepsilon_{\mathcal{C}}(\vec{r})$ in the cube sites. 
As outlined in Fig.~\ref{fig:dir_to_ord} using a 2D section of the full $\mathcal{L}_3(1)$ lattice for convenience, we restore the order in the systems along the following three steps:
(i) we render the disorder on the cube sites binary $\varepsilon_\mathcal{C}(\vec{r}) =  \pm \mu$ -- {\it i.e.} yielding a randomized checkerboard potential of strength $\mu$ [panel (a)]; 
(ii) we reorder the randomized checkerboard $\varepsilon_\mathcal{C}(\vec{r}) = (-1)^{\bar{\vec{r}}}\mu$ where $\bar{\vec{r}} = x+y+z$ -- {\it i.e.} yielding an ordered checkerboard potential of strength $\mu$ [panel (b)]; and
(iii) we realign all the checkerboard values yielding a constant potential shift only on a sub-lattice [panel (c)]. 
In order to mimic experimentally relevant set-ups, each of these three cases will be studied including an additional weak uncorrelated impurities $\delta_\mathcal{C}(\vec{r})\in \left[-\frac{W}{2}, \frac{W}{2}\right]$ of strength $W$ in the cube potential $\varepsilon_\mathcal{C}(\vec{r}) + \delta_\mathcal{C}(\vec{r})$ as schematically represented in panels (d-f) of Fig.~\ref{fig:dir_to_ord}. 


\section{Method}
\label{sec:method}

In the following, we report on results obtained by numerically diagonalizing a finite $\mathcal{L}_3(1)$ lattice, Eq.~\eqref{Equ:def1}, with periodic boundary conditions and $M$ unit cells per side, yielding a total of $L = 4 M^3$ lattice sites. We are interested in the complete set of energy eigenvalues and corresponding eigenstates $\Psi_l$, $l = 1, \ldots, 4 M^3$ and employ the standard LAPACK \cite{LAPACK2012} sub-routine {\tt DSYEV} to generate these. 


In the aforementioned Wannier basis, we can write
\begin{equation}
\Psi_l = \sum_{\vec{r}\in\mathcal{L}_3(1)} \psi_l(\vec{r}) |\vec{r} \rangle ,
\end{equation}
where $\psi_l(\vec{r})$ denotes the amplitude of eigenstate $\Psi_l$, with eigenenergy $E_l$, projected onto lattice site $\vec{r}$ with tight-binding Wannier state $| \vec{r} \rangle$. 
The \emph{participation ratio} is then defined as 
\begin{equation}
    P(E_l)= \frac{1}{L \sum_{\vec{r}\in\mathcal{L}_3(1)} |\psi_l{(\vec{r})}|^4}.
\end{equation}
In order to familiarize the reader with this perhaps somewhat unusual version of the ``participation'' concept \cite{Mirlin2000}, let us briefly consider two limiting cases. Without any disorder or onsite potential variation, we can assume a Bloch-like structure of the wave function such that $\psi_l{(\vec{r})} \propto 1/\sqrt{L}$ due to normalization. Then $P(E_l)= 1/ ( L \sum_{\vec{r}\in\mathcal{L}_3(1)} 1/L^2 ) = 1$. On the other hand, when only a single site $\vec{r}_0$ has a non-zero wave function amplitude, such as in a perfectly localized state, we have $P(E_l)= 1/ ( L \sum_{\vec{r}\in\mathcal{L}_3(1)} \delta(\vec{r},\vec{r}_0)) = 1/L$. The participation ratio hence is $\simeq 1$ for extended states while it is $\simeq 0$ for large enough $L$; it measures the ratio of lattice sites which have a non-zero wavefunction amplitude. In the following, it shall serve as our measure of the strength of localization in $\mathcal{L}_3(1)$.


We are also interested in the distribution of the wave function intensity across $\mathcal{L}_3(1)$. Trivially, 
$Q(E_l) \equiv \sum_{\vec{r}\in\mathcal{L}_3(1)} |\psi_l(\vec{r})|^2 = 1$ by normalization. More useful in the present case is to split the sum into sites in $\mathcal{C}$ and those in $\mathcal{L}$. We then define
\begin{equation}
    Q_{\mathcal{C}}(E_l)  =  \sum_{\vec{r}\in\mathcal{C}} |\psi_l(\vec{r})|^2  , \quad
    Q_{\mathcal{L}}(E_l)  =  \sum_{\vec{r}\in\mathcal{L}} |\psi_l(\vec{r})|^2  ,
\end{equation}
while always $Q_{\mathcal{C}}(E_l) + Q_{\mathcal{L}}(E_l) = 1$. 
Similarly, we can define
\begin{equation}
    P_{\mathcal{C}}(E_l)= \frac{1}{|\mathcal{C}| \sum_{\vec{r}\in\mathcal{C}} |\psi_l{(\vec{r})}|^4} , \quad
    P_{\mathcal{L}}(E_l)= \frac{1}{|\mathcal{L}| \sum_{\vec{r}\in\mathcal{L}} |\psi_l{(\vec{r})}|^4} ,
\end{equation}
where $|\mathcal{C}|$ and $|\mathcal{L}|$ give the number of sites in $\mathcal{C}$ and $\mathcal{L}$, respectively.
The quantities $Q_{\mathcal{C}}(E_l)$ and $Q_{\mathcal{L}}(E_l)$ then measure how much wave function intensity the state $\Psi_l$ at energy $E_l$ retains on cube ($\mathcal{C}$) and Lieb ($\mathcal{L}$) sites while $P_{\mathcal{C}}(E_l)$ and $P_{\mathcal{L}}(E_l)$ indicate how extended the state is across $\mathcal{C}$ and $\mathcal{L}$, respectively. 
 
\section{Results}
\label{sec:results}

In this section we analyze the complete eigenspectrum and eigenstates of $\mathcal{L}_3(1)$ by computing projected probabilities $Q_{\mathcal{C}}$, $Q_{\mathcal{L}}$ and the participation numbers $P_{\mathcal{C}}$, $P_{\mathcal{L}}$ for three choices of spatial onsite energy distributions $\varepsilon_\mathcal{C}(\vec{r})$ shown schematically in Fig.~\ref{fig:dir_to_ord}(a-c).
%
We perform our tests for $40$ equi-spaced strengths $\mu$ of the cube-site potential $\varepsilon_\mathcal{C}(\vec{r})$ from $\mu=0.0, 0.25, 0.5, \ldots, 10$, and three different strengths of perturbation $\delta_{\vec{r}}^{(c)}$, see Fig.\ \ref{fig:dir_to_ord}(d-f), namely, $W=1$, $2$ and $5$. 
In each case, we diagonalize $\mathcal{L}_3(1)$ with $L=4\times 10^3$ sites.
The  values of $P_\mathcal{C}$, $P_\mathcal{L}$  and $Q_\mathcal{C}$, $Q_\mathcal{L}$ are averaged over $R=64$ realizations, and each point in Fig.~\ref{fig:dir_to_ord} corresponds to an average over $\sim 640$ energies within an energy window of width $0.05$. 
We have tested that increasing $L$ to $4\times 20^3$ and $R$ to $1000$ does not change the results significantly, but just makes the analysis numerically more cumbersome.
Let us note that we account $L/2$ eigenstates only from each realization, as we exclude from the plots the computed eigenstates with $|E| \leq 10^{-4}$, i.e.\ the degenerate CLSs at $E=0$.

\subsection{Randomized checkerboard potential}
\label{sec:random-checkerboard}

\begin{figure}[tb]
    \centering
    \hspace*{0.14\textwidth}  $W=1$ \hfill $W=2$ \hfill $W=5$ \hspace*{0.10\textwidth} \\[1ex]
    (a)\includegraphics[width=0.29\textwidth]{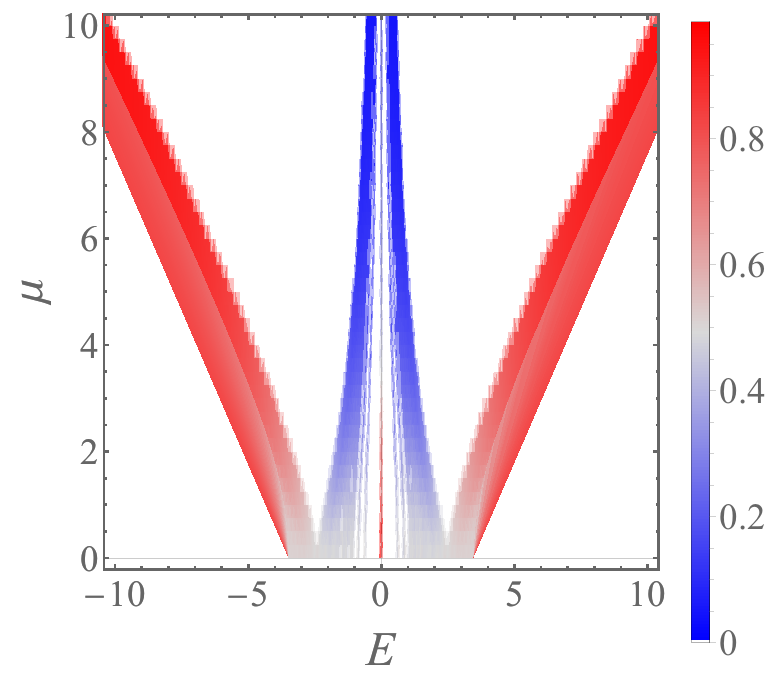}
    (b)\includegraphics[width=0.29\textwidth]{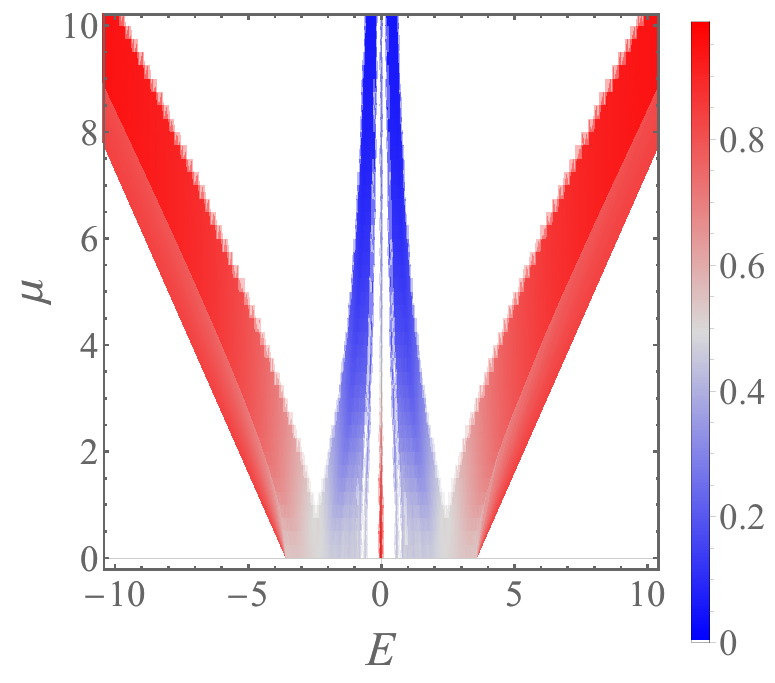}
    (c)\includegraphics[width=0.29\textwidth]{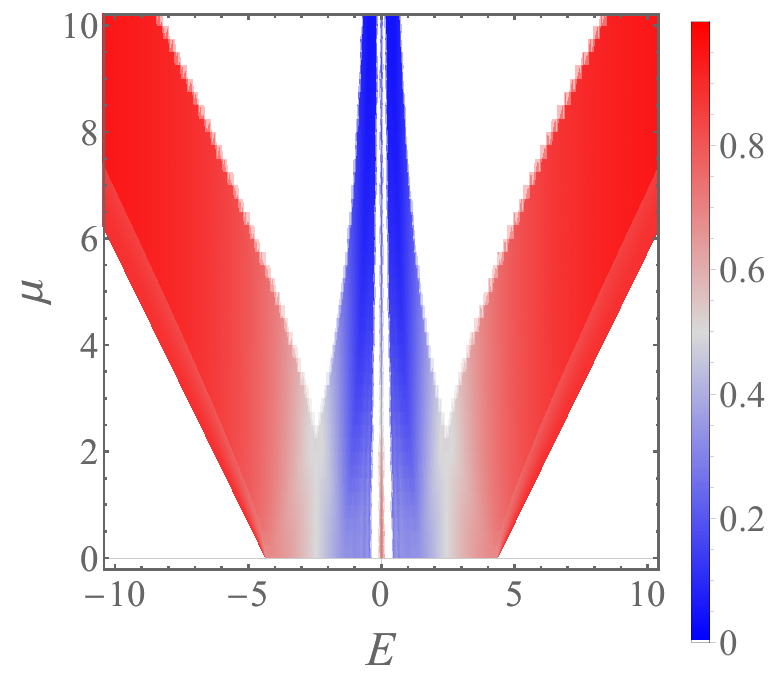}\\
    (d)\includegraphics[width=0.29\textwidth]{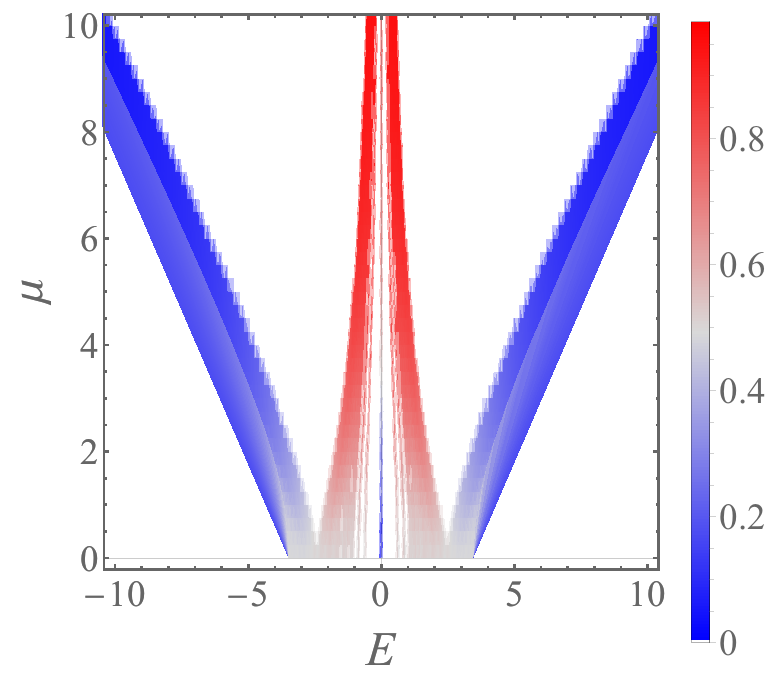}
    (e)\includegraphics[width=0.29\textwidth]{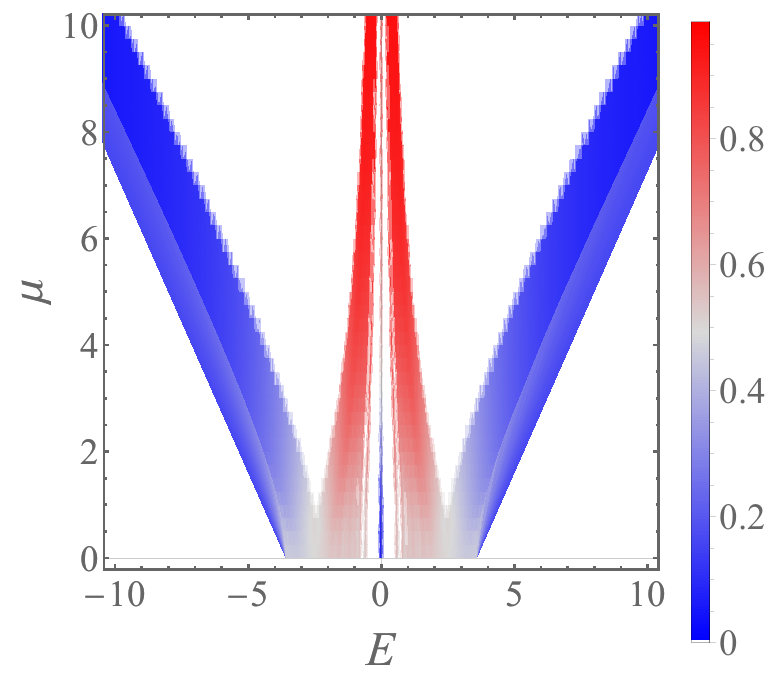}
    (f)\includegraphics[width=0.29\textwidth]{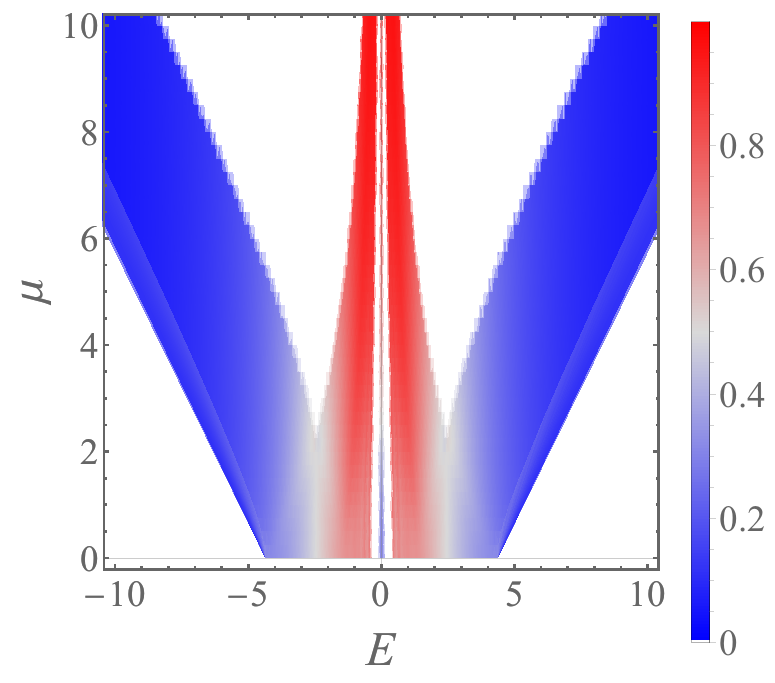}  %
    \caption{
    Projected probabilities $Q_{\mathcal{C}}$ of the \emph{random} checker board and their energy $E$ and $\mu$ dependence for disorder strength (a) $W=1$, (b) $2$ and (c) $5$ at $L=4000$, averaged over $R=64$ disorder realizations.
    Panels (d-f) are similar to (a-c) but for the $Q_{\mathcal{L}}$, i.e.\ on Lieb sites.
    The colours from blue to grey to red denote the range of values from $0$ to $1$ in each panel. We emphasise that white denotes the absence of states.
    }
    \label{fig:probs_l31_CP2}
\end{figure}

\begin{figure}[tb]
    \centering
    \hspace*{0.14\textwidth}  $W=1$ \hfill $W=2$ \hfill $W=5$ \hspace*{0.10\textwidth} \\[1ex]
    (a)\includegraphics[width=0.29\textwidth]{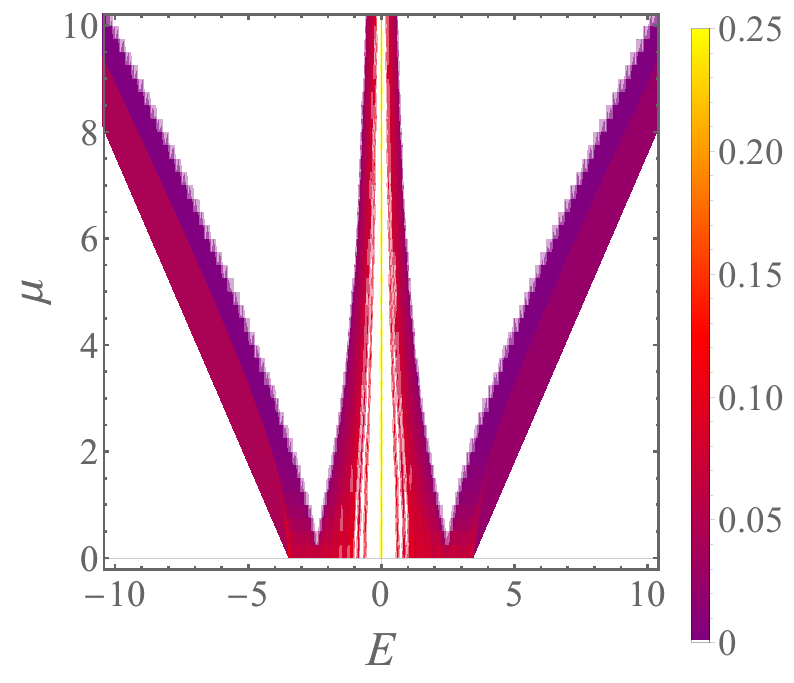}
    (b)\includegraphics[width=0.29\textwidth]{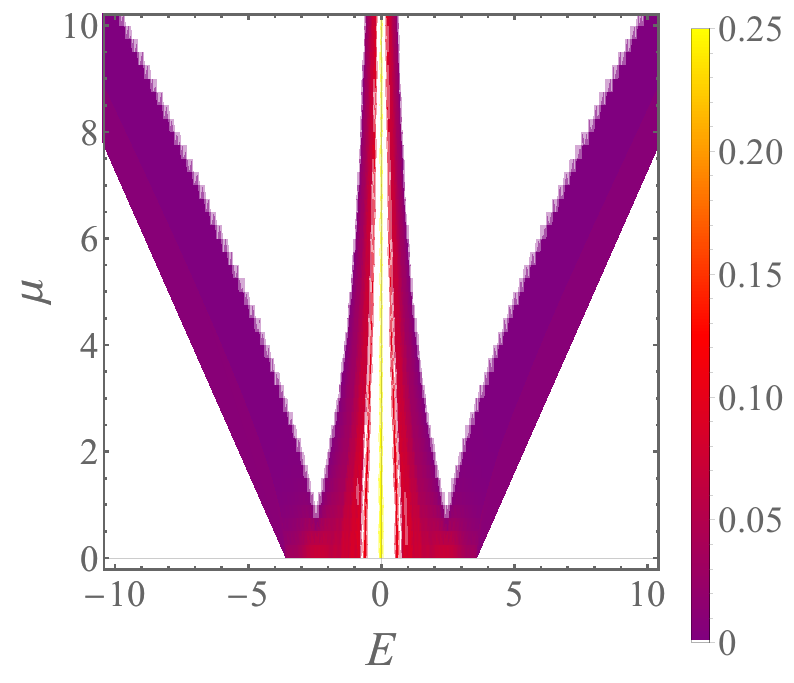}
    (c)\includegraphics[width=0.29\textwidth]{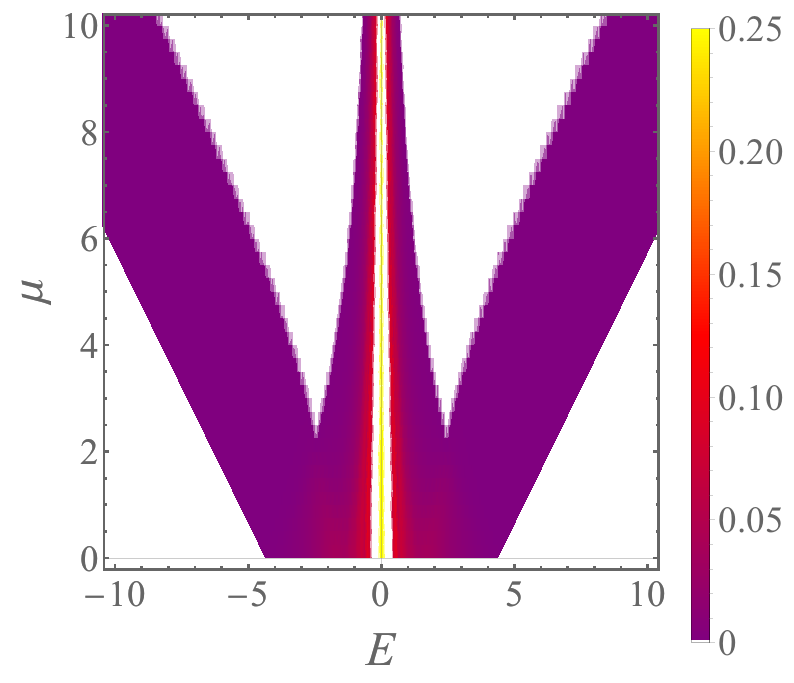}
    (d)\includegraphics[width=0.29\textwidth]{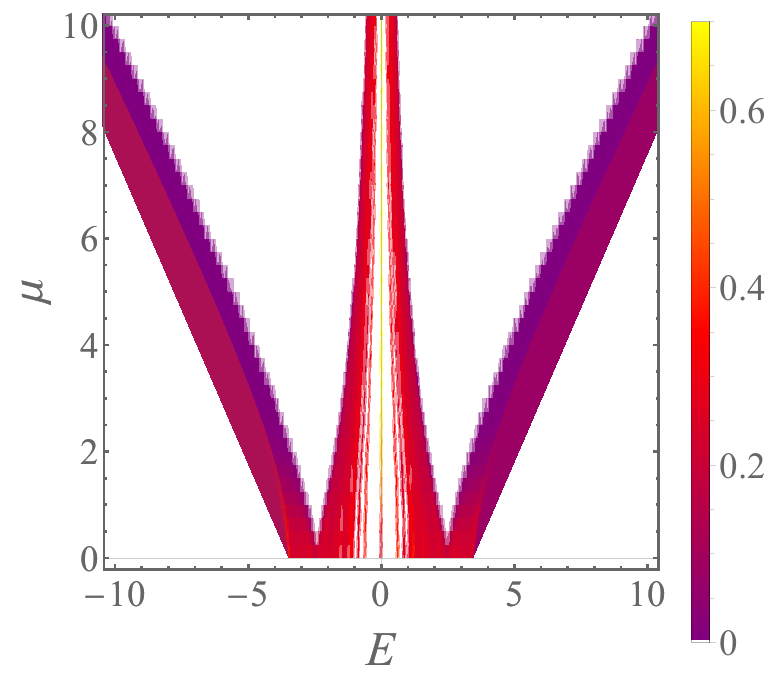}
    (e)\includegraphics[width=0.29\textwidth]{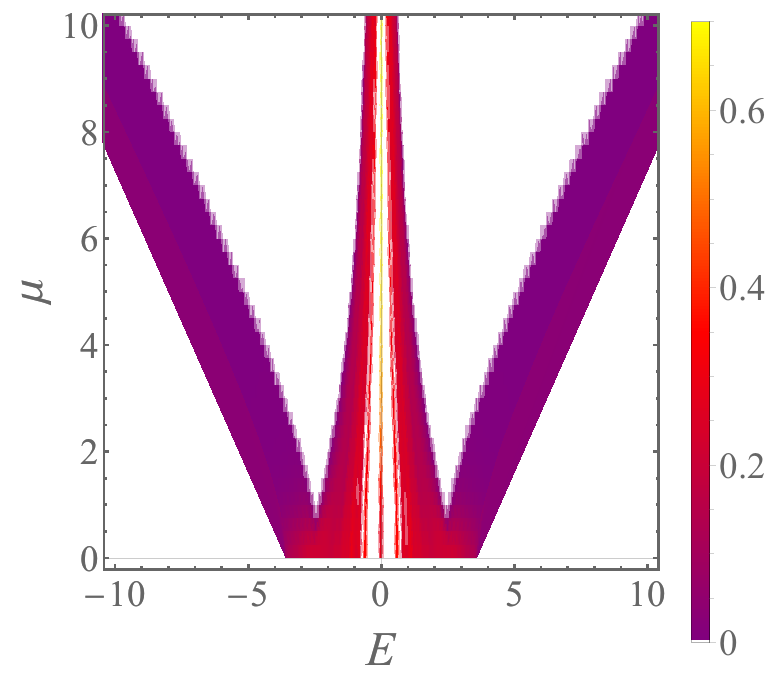}
    (f)\includegraphics[width=0.29\textwidth]{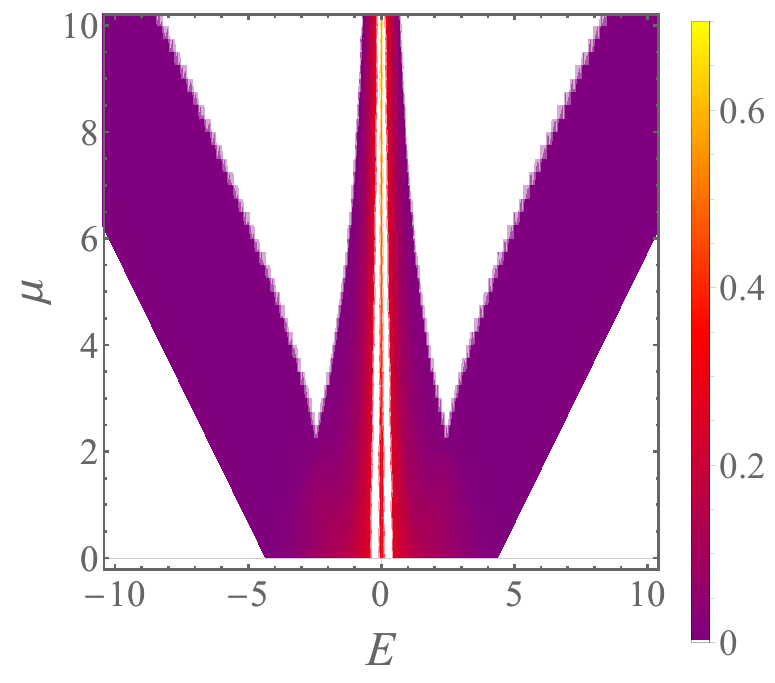}  
    \caption{
    Participation numbers $P_{\mathcal{C}}$ of the \emph{random} checker board and their energy $E$ and $\mu$ dependence for disorder strength (a) $W=1$, (b) $2$ and (c) $5$ for $L$ and sample number $R$ as in Fig.\ \ref{fig:probs_l31_CP2}.
    Panels (d-f) are similar to (a-c) but for the $P_{\mathcal{L}}$ of Lieb sites. Color ranges between $P_{\mathcal{C}}$ and $P_{\mathcal{L}}$ are different. As in Fig.\ \ref{fig:probs_l31_CP2}, a white color marks the absence of states.
    }
    \label{fig:Pn_l31_CP2}
\end{figure}

The disordered checkerboard is shown in Fig.~\ref{fig:dir_to_ord}(a). 
This case is a direct descent of the one previously considered in Ref.~\cite{Liu2022}, where an uncorrelated cube potential disorder over a continuous interval $\varepsilon_\mathcal{C}(\vec{r})\in [-\mu,+\mu]$ was considered. Here we instead study the binary disorder potential with $\varepsilon_\mathcal{C}(\vec{r}) = \pm\mu$.  
The potential is additionally perturbed by a weak uncorrelated potential $ \delta_{\vec{r}}^{(c)}$ as indicated in Fig.~\ref{fig:dir_to_ord}(d).
%
%
We note that binary disorder already has its own research literature in the context of Anderson localization \cite{Economou1988,PLYUSHCHAY2003,Karmann2006} and is often used as a convenient starting point for mathematical proofs of localization \cite{Stollmann2001CaughtDisorder,Imbrie2021}.

\subsubsection{Projected probabilities and spectral gaps}

We begin by discussing the projected probabilities $Q_\mathcal{C,L}$ of the eigenstates as function of the potential strength $\mu$. 
In Fig.~\ref{fig:probs_l31_CP2} we show the projection $Q_{\mathcal{C}}$ on the cube sub-lattice $\mathcal{C}$ [panels (a-c)] and the probability $Q_{\mathcal{L}}$ on the Lieb sub-lattice $\mathcal{L}$ [panels (d-f)]  for three different values of the perturbation strength $W$ as function of (eigen) energies $E_l$ and $\mu$.
All cases show the emergence of gaps separating two families of states characterized by substantially different behaviours of the projected probabilities $Q_{\mathcal{C}}$ and $Q_{\mathcal{L}}$. 

For instance, with perturbation strength $W=1$ as shown in Fig.~\ref{fig:probs_l31_CP2}, panels (a+d), we observe a group of states with a low value of $Q_{\mathcal{C}} \lesssim 0.2$, cp.\ panel (a), close to the macroscopic degeneracy energy $E=0$, while in the same energy regime, $Q_{\mathcal{L}}$ is much larger, cp.\ panel (d). This indicates that these states mostly occupy the clean Lieb sub-lattice $\mathcal{L}$. 
Such states behave thus similar to the situation with uncorrelated disorder in the cube lattice studied in Ref.~\cite{Liu2022}. 
Meanwhile, we observe another groups of states which mostly occupy $\mathcal{C}$, with high value of $0.8 \lesssim Q_{\mathcal{C}} \lesssim 1$, see Fig.~\ref{fig:probs_l31_CP2}(a), whose energies grow in absolute values proportionally with $\mu$. 
As expected, these observations are again corroborated by $Q_{\mathcal{L}}$, cp.\ panel (d)\footnote{We note that as the projected probabilities $Q_{\mathcal{C}}$ and $Q_{\mathcal{L}}$ are complementary to each other, we will in other plots only show $Q_{\mathcal{L}}$.}.

The gaps separating the two energy groups are broadening up as $\mu$ grows, while at $E \approx 0$, a minor fraction of states, which mostly occupy $\mathcal{C}$, appears -- seemingly gapped from the surrounding states located in $\mathcal{L}$ .  
This group consists of those with one-state-per-realization at energy $E\sim 10^{-2}$ (at this $L=4000$ system size) which are due to an accidental degeneracy following the self-averaging of the discretized potential $\langle \varepsilon_\mathcal{C}(\vec{r}) \rangle_{L\rightarrow \infty} \rightarrow 0$ similar to what observed in Ref.~\cite{Liu2022}. 
Interestingly, we notice that these accidentally degenerate states turn from populating $\mathcal{C}$ for low $\mu$ to $\mathcal{L}$ for increasing $\mu$ -- as it becomes energetically favorable to occupy the latter lattice as $\mu$ increases. 

Stronger values of the perturbation strengths $W=2$ and $W=5$ -- shown respectively in Fig.~\ref{fig:probs_l31_CP2}(b+e) and (c+f) -- reconfirm the above discussed picture for $W=1$, and retain the existence of these two distinguished groups of states.  
In both cases, we observe that increasing the perturbation strength $W$ broadens the spectral width of both families of states -- yielding a gap between the two families opening at larger $\mu$. 
Hence, the random checkerboard separates states confined to $\mathcal{L}$ with low energy $E \lesssim 2.5$ from those confined on $\mathcal{C}$ with $E \gtrsim 2.5$.

\subsubsection{Participation number and localization}

We now wish to ascertain the spatial extent of the two group of states identified above, localized either on $\mathcal{C}$ or $\mathcal{L}$ sites.
In Fig.~\ref{fig:Pn_l31_CP2} we show, respectively, the participation numbers $P_{\mathcal{C,L}}$ for the three different perturbation strengths $W=1$, $2$ and $5$. The panel arrangement is analogous as in Fig.~\ref{fig:probs_l31_CP2}. 
In Fig.~\ref{fig:Pn_l31_CP2}(a) we observe that for $W=1$ the $P_{\mathcal{C}}$ values are low at $\lesssim 0.05$ for the higher energy states, while they reaches substantially larger values of $\gtrsim 0.2$ for the lower energy states. The highest values are reached for energies very close to the macroscopic degeneracy $E=0$.
Similarly this follows from the simulations of the participation number $P_{\mathcal{L}}$ in the Lieb sub-lattice shown in Fig.~\ref{fig:Pn_l31_CP2}(d). 
Likewise, these conclusions hold also for stronger values of the additional perturbation $W$ -- as shown in Fig.~\ref{fig:Pn_l31_CP2}(b,e) and (c,f). Hence, in comparison, the low energy states are spatially more extended while the high energy states show signatures of more localization, and these features are robust to the presence of perturbations. 

These results are in agreement with those 
obtained in Ref.~\cite{Liu2020a}. Indeed,  uncorrelated cube potential disorder over a continuous interval $[-\mu,+\mu]$ induces the coexistence of low energy extended states over the Lieb sub-lattice with localized states over the cube sub-lattice. 
However, in the present case the binary discretized disorder yields that these two distinct families of states are separated by band gaps. Such gaps are tunable, as their width can be controlled by tuning the potential strength $\mu$. Furthermore, they are robust to perturbation. 

\subsection{Ordered checkerboard potential}
\label{sec:ordered-checkerboard}

\begin{figure}[tb]
    \centering
    \hspace*{0.14\textwidth}  $W=1$ \hfill $W=2$ \hfill $W=5$ \hspace*{0.10\textwidth} \\[1ex]
    (a)\includegraphics[width=0.29\textwidth]{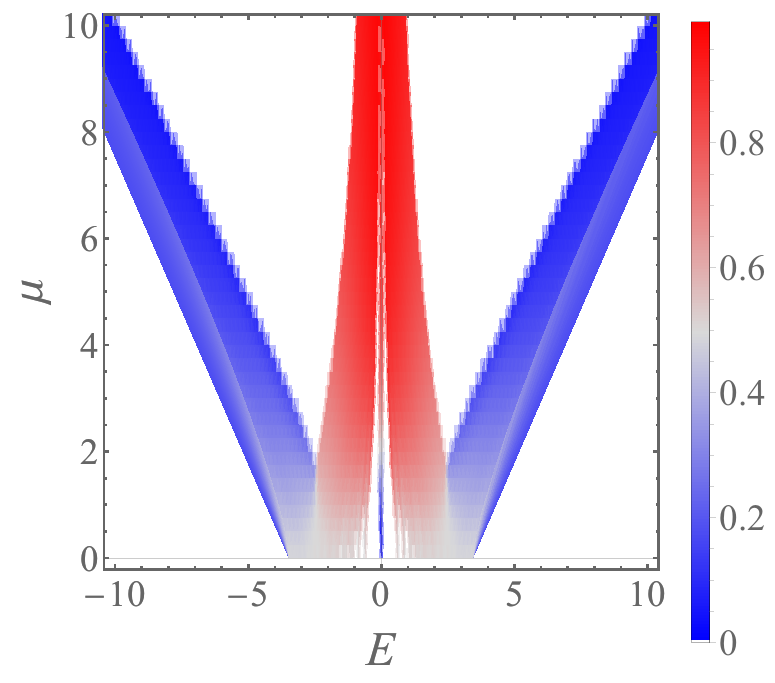}
    (b)\includegraphics[width=0.29\textwidth]{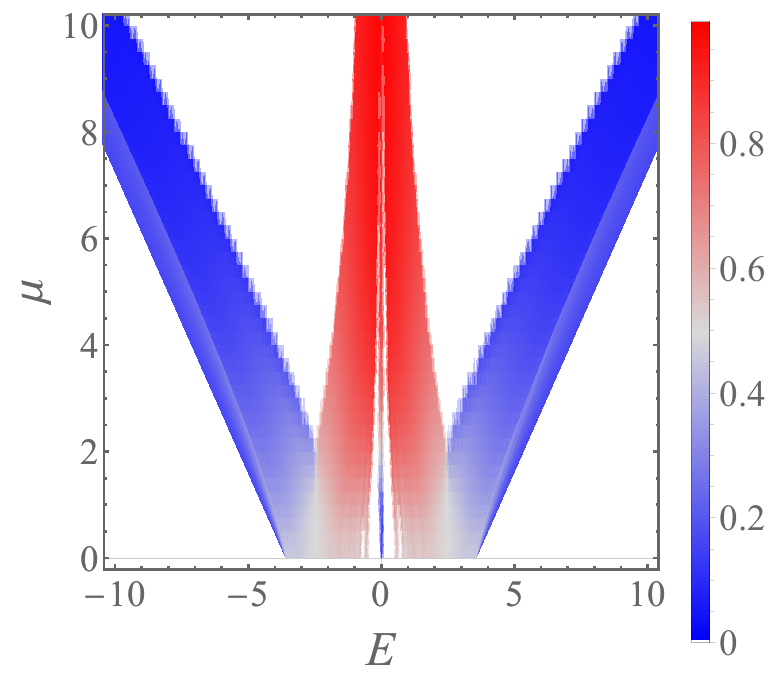}
    (c)\includegraphics[width=0.29\textwidth]{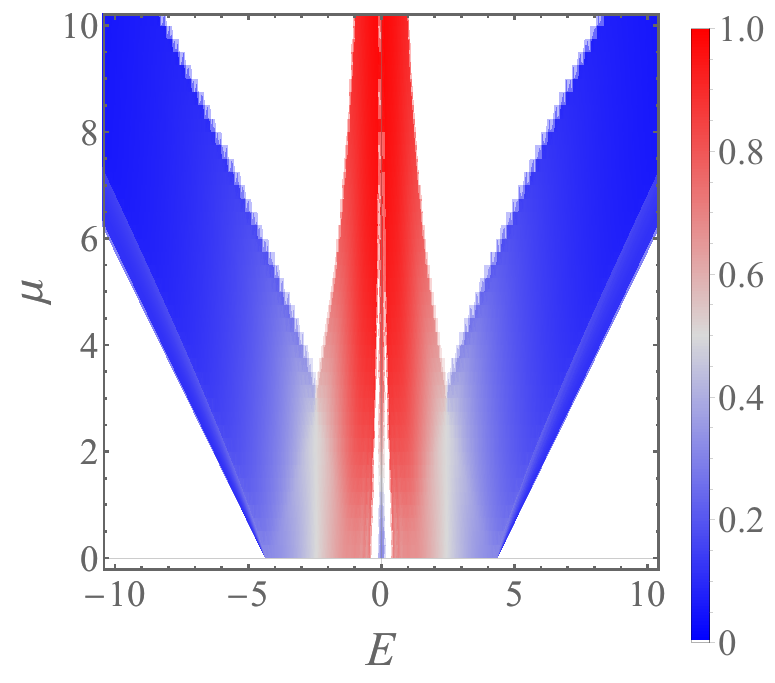}    
    \caption{Projected probabilities $Q_{\mathcal{L}}$ of the \emph{ordered} checker board and their energy $E$ and $\mu$ dependence for disorder strength (a) $W=1$, (b) $2$ and (c) $5$ 
    as in Fig.\ \ref{fig:probs_l31_CP2}.
    Due to the normalization condition, $Q_{\mathcal{C}}(E_l) + Q_{\mathcal{L}}(E_l) = 1$, the results for $Q_{\mathcal{C}}(E_l)$ simply switch the colours as in Fig.\ \ref{fig:probs_l31_CP2} and are hence not shown.
    }
    \label{fig:probs_l31_CP1}
\end{figure}

\begin{figure}[tb]
    \centering
    \hspace*{0.14\textwidth}  $W=1$ \hfill $W=2$ \hfill $W=5$ \hspace*{0.10\textwidth} \\[1ex]
    (a)\includegraphics[width=0.29\textwidth]{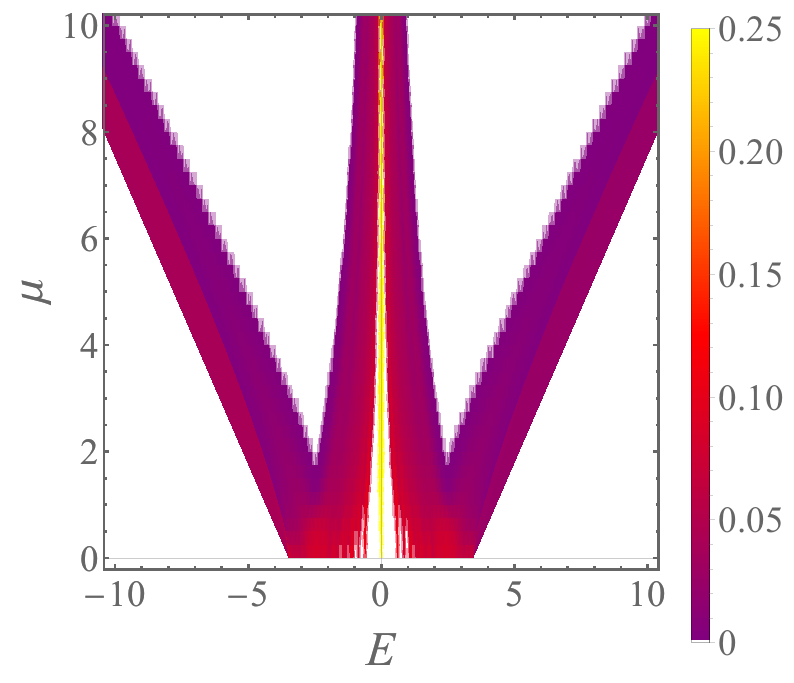}
    (b)\includegraphics[width=0.29\textwidth]{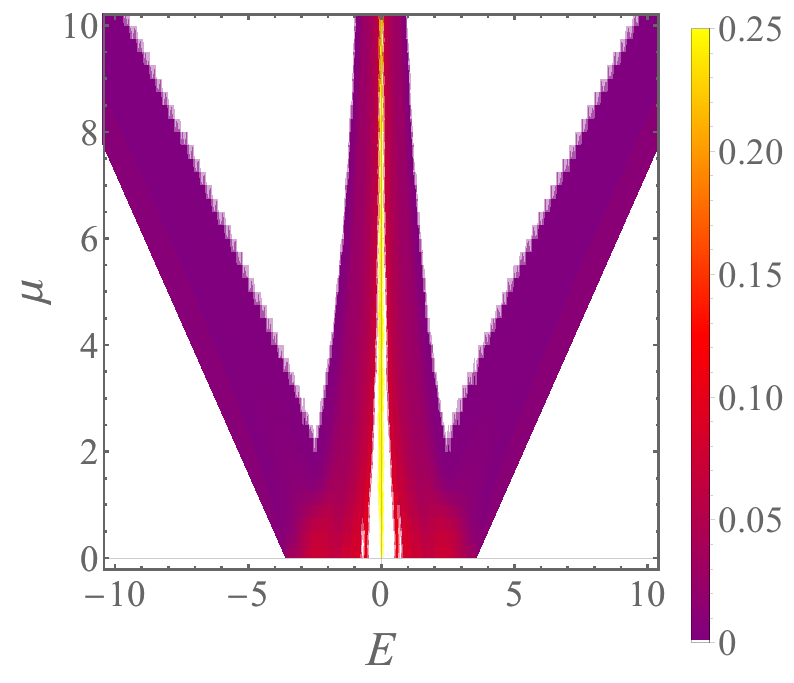}
    (c)\includegraphics[width=0.29\textwidth]{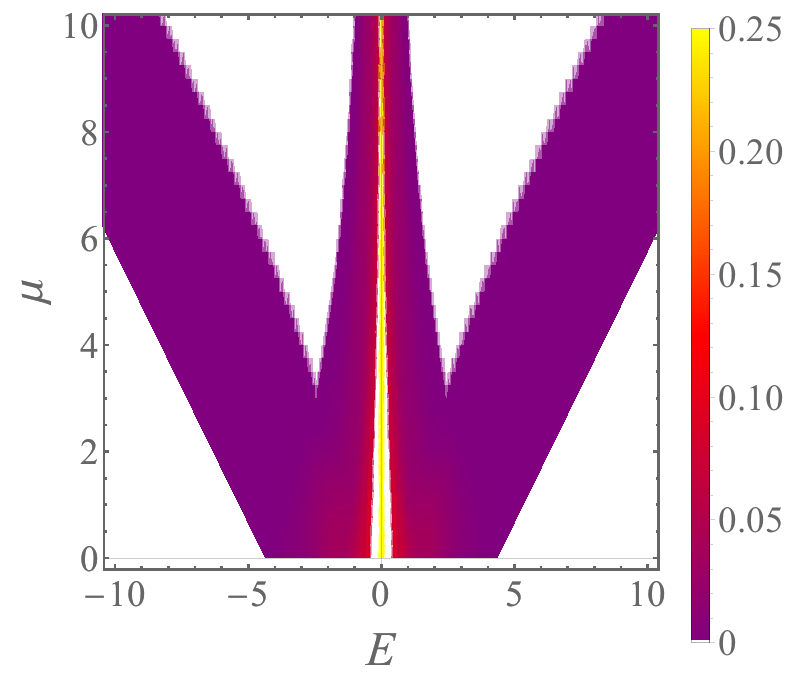}
    (d)\includegraphics[width=0.29\textwidth]{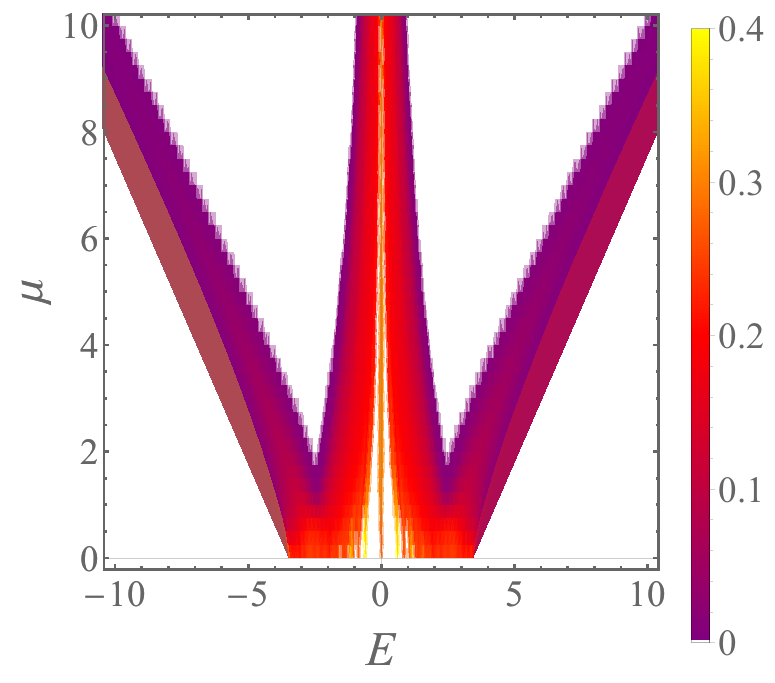}
    (e)\includegraphics[width=0.29\textwidth]{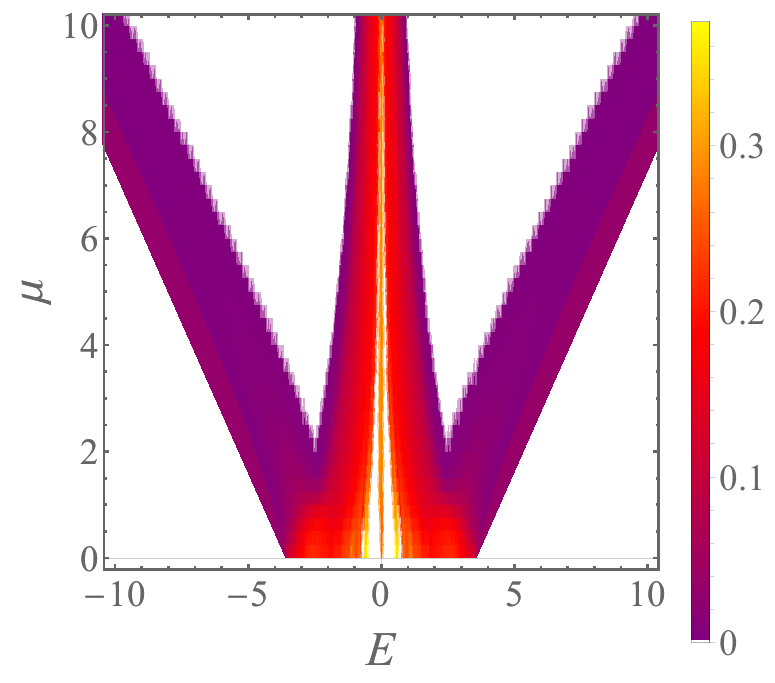}
    (f)\includegraphics[width=0.29\textwidth]{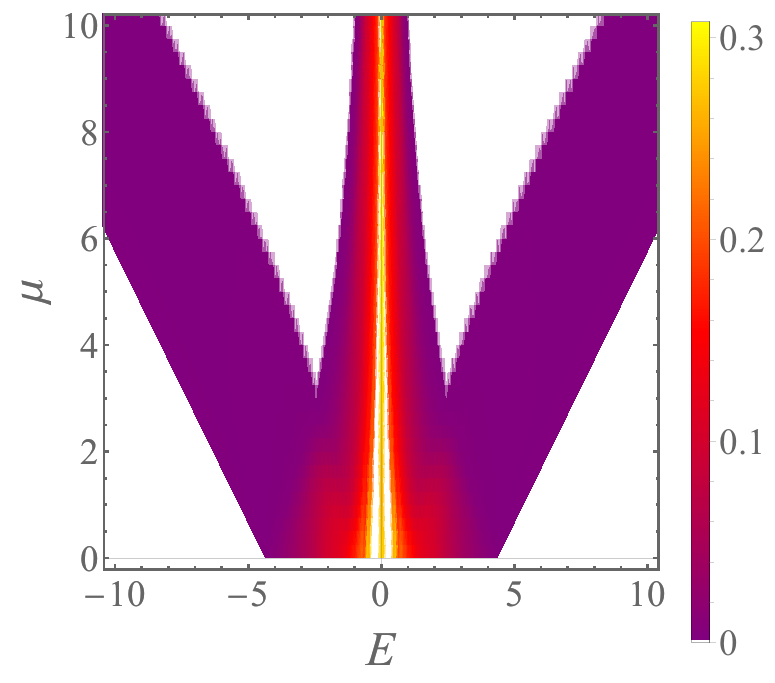}    
    \caption{
    Participation numbers $P_{\mathcal{C}}$ of the \emph{random} checker board and their energy $E$ and $\mu$ dependence for disorder strength (a) $W=1$, (b) $2$ and (c) $5$ for $L$ and $R$ as in Fig.\ \ref{fig:probs_l31_CP2}.
    Panels (d-f) are similar to (a-c) but for the $P_{\mathcal{L}}$ of Lieb sites. Color ranges between $P_{\mathcal{C}}$ and $P_{\mathcal{L}}$ are different. As in Fig.\ \ref{fig:probs_l31_CP2}, a white color marks the absence of states.
    }
    \label{fig:Pn_l31_CP1}
\end{figure}

Reordering the discretized uncorrelated disorder $\varepsilon_\mathcal{C}(\vec{r})\in\{-\mu,+\mu\}$ in the cube sites $\mathcal{C}$ -- {\it i.e.} 
switching to an ordered checkerboard  $\varepsilon_\mathcal{C}(\vec{r}) = (-1)^{x+y+z}\mu$ 
shown in Fig.~\ref{fig:dir_to_ord}(b) -- restores translation invariance in the system, which therefore result in extended states and dispersive Bloch bands. 
In this case, a minimal unit cell is obtained by gluing together eight of the neighboring four-site unit cells  highlighted with gray cubes in Fig.~\ref{fig:Lieb_schematic}(a), in order to form a larger cube. 
This yields $32$ bands for the lattice $\mathcal{L}_3(1)$ -- $16$ of which are flat at $E=0$ regardless on the size of $\mu\neq 0$, while the remaining $16$ dispersive bands are symmetrically split in eight positive and eight negative. 
The checkerboard potential induces band gaps between the dispersive and the flat bands \footnote{The analytical expressions of the resulting dispersive bands for $\mu\neq 0$ are too cumbersome to be reported here.}, which {\it e.g.} at the $R = (\pi, \pi, \pi)$ point of the spectrum, the gap width increasing linearly with $\mu$. 
In our tests of the robustness of the spatial features of the eigenstates, we perturb the ordered checkerboard with an uncorrelated potential $ \delta_{\vec{r}}^{(c)}$, as presented in Fig.~\ref{fig:dir_to_ord}(e). 

\subsubsection{Projected probabilities and spectral gaps}

In Fig.~\ref{fig:probs_l31_CP1} (a-c) we show that for each $W=1$, $2$ and $5$ we observe that as $\mu$ grows again two families of states emerge which are gapped away from each other. 
These families  -- as for the disordered checkerboard in Fig.~\ref{fig:probs_l31_CP2} -- consist of low energy states which mostly sit in the Lieb sub-lattice $\mathcal{L}$ ({\it i.e.} high value of $Q_{\mathcal{L}}$) and high energy states (in absolute value) which mostly sit in the cube sub-lattice $\mathcal{C}$ ({\it i.e.} high value of $Q_{\mathcal{C}}$ close to one).  
These families in the energy spectrum are separated by the gaps which broaden up as $\mu$ grows. 
Note that, as this potential is also self-averaging  $\langle \varepsilon_\mathcal{C}(\vec{r}) \rangle_{L\rightarrow \infty} \rightarrow 0$, accidental degenerate states at $E\approx 10^{-2}$ appear. 
Furthermore, alike the former case, the spectral widths of both families of states increase as the strength $W$ of the additional perturbation grows -- indicating that also in this case the spectral gaps are robust against onsite perturbation. 

\subsubsection{Participation numbers and localization}

The participation number $P_\mathcal{C}$ on $\mathcal{C}$ shown in Fig.~\ref{fig:Pn_l31_CP1}(a-c) reveal that 
the high energy states have low values ({\it i.e.} indicating localization) 
while the low energy states have comparatively higher values,{\it i.e.}, indicating lesser localized states. 
We note that, the values of $P_\mathcal{C}$ also decrease as $W$ grows. 
This contrast in behaviour of the participation number between the two families of high and low energy states can be seen even more clearly in Fig.~\ref{fig:Pn_l31_CP1}(d-f)
for the projected $P_\mathcal{L}$. 

The mechanism of separation in two distinct and gapped families of eigenstates in this case relies on the band gap initiated by the ordered checkerboard potential. 
Interestingly, this potentials maintain the main features seen in the previous case with the random checkerboard. 
Hence, this ordered checkerboard potential with the additional weak perturbation (unavoidable in experimental setups) can be an easier way to achieve a split in localized high energy states concentrated in one sub-lattice, from the low energy delocalized eigenstates focused instead in the complementary sub-lattice.  

\subsection{Constant-shift potential}
\label{sec:constant-shift}

\begin{figure}[tb]
    \centering
    \hspace*{0.14\textwidth}  $W=1$ \hfill $W=2$ \hfill $W=5$ \hspace*{0.10\textwidth} \\[1ex]
    (a)\includegraphics[width=0.29\textwidth]{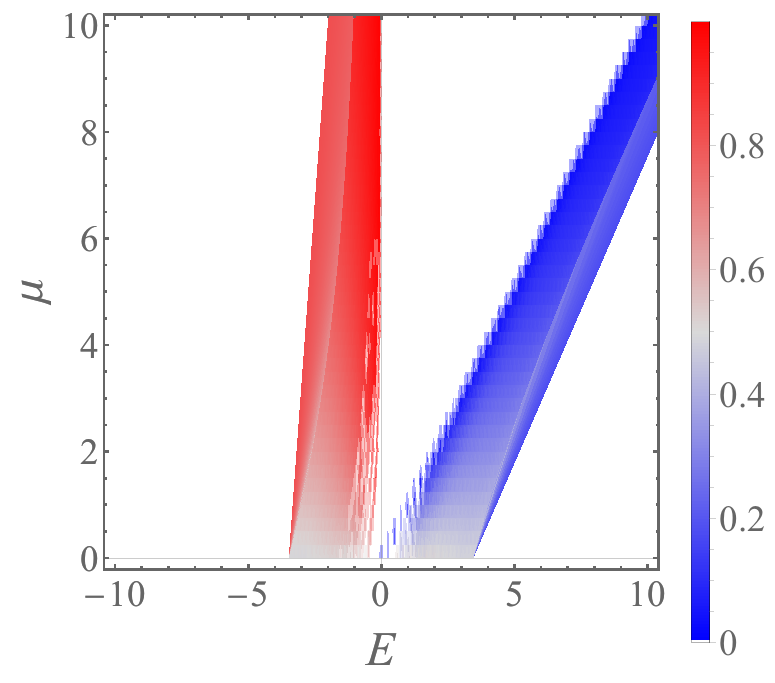}
    (b)\includegraphics[width=0.29\textwidth]{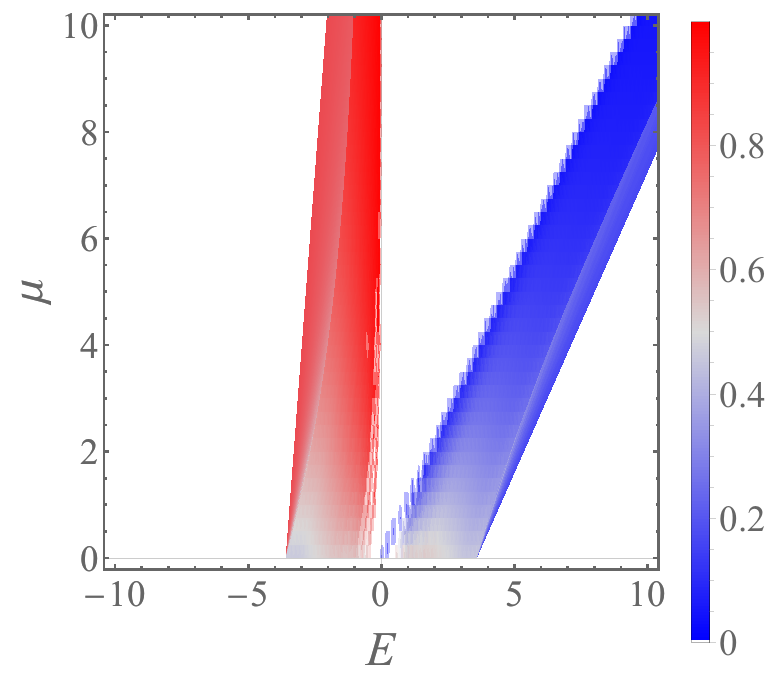}
    (c)\includegraphics[width=0.29\textwidth]{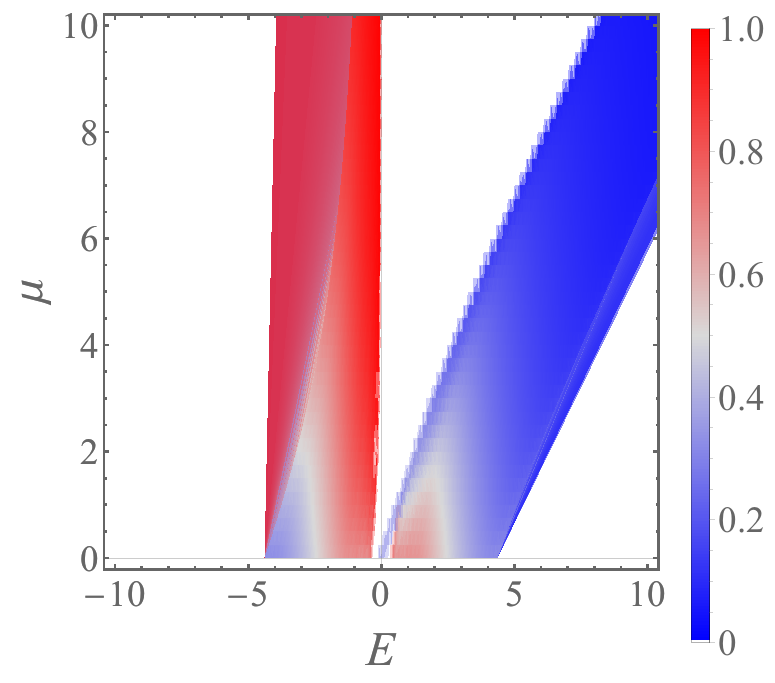}    
    \caption{Projected probabilities $Q_{\mathcal{L}}$ for \emph{constant shift} potential and their ($E$, $\mu$) dependence with disorder strength (a) $W=1$, (b) $2$ and (c) $5$.
    System size, number of configurations and colours/legends as in Fig.\ \ref{fig:probs_l31_CP2}.
    }
    \label{fig:probs_l31_CP0}
\end{figure}

\begin{figure}[tb]
    \centering
    \hspace*{0.14\textwidth}  $W=1$ \hfill $W=2$ \hfill $W=5$ \hspace*{0.10\textwidth} \\[1ex]
    (a)\includegraphics[width=0.29\textwidth]{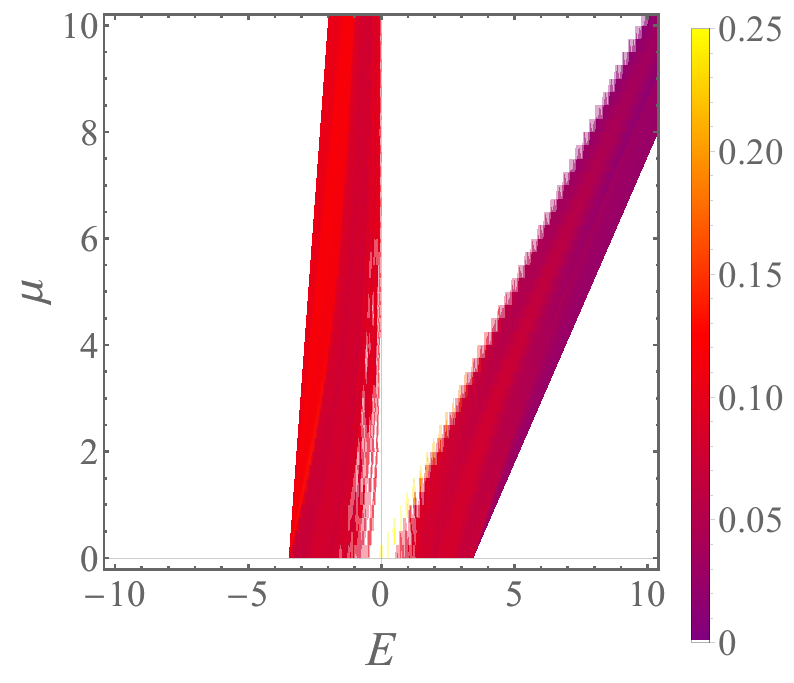}
    (b)\includegraphics[width=0.29\textwidth]{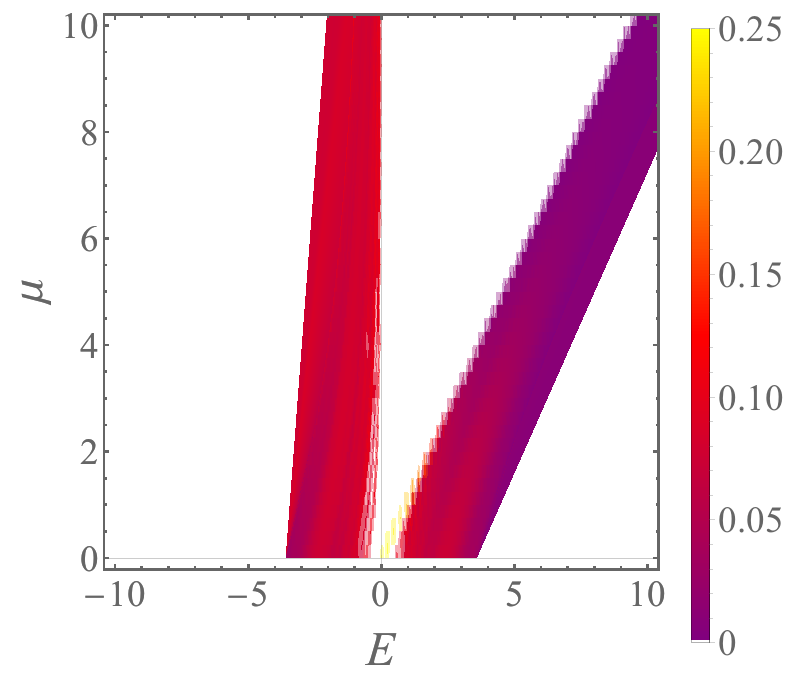}
    (c)\includegraphics[width=0.29\textwidth]{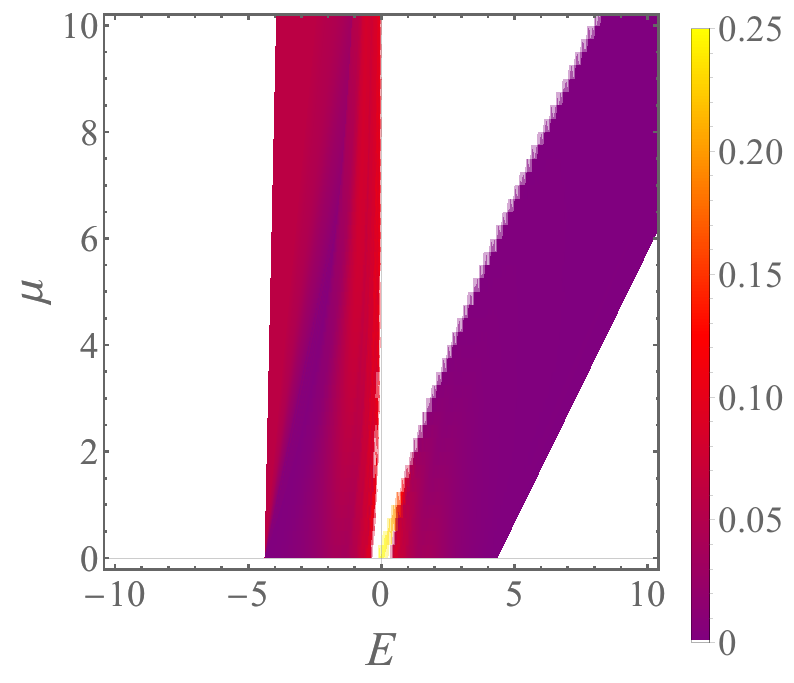}
    (d)\includegraphics[width=0.29\textwidth]{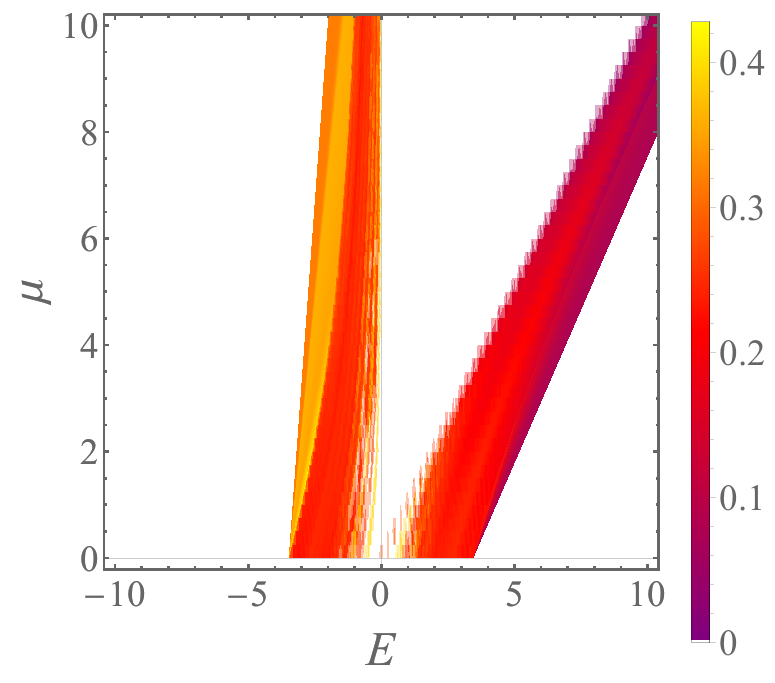}
    (e)\includegraphics[width=0.29\textwidth]{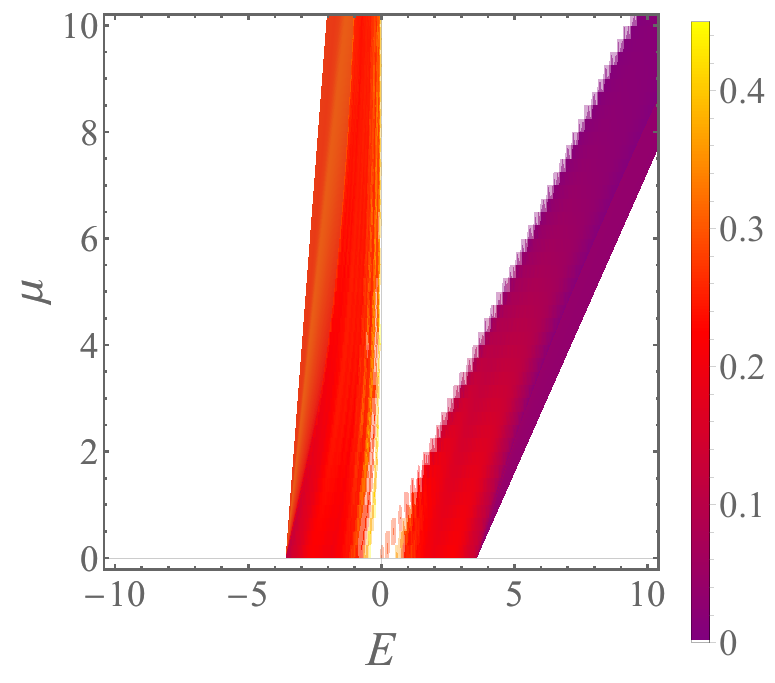}
    (f)\includegraphics[width=0.29\textwidth]{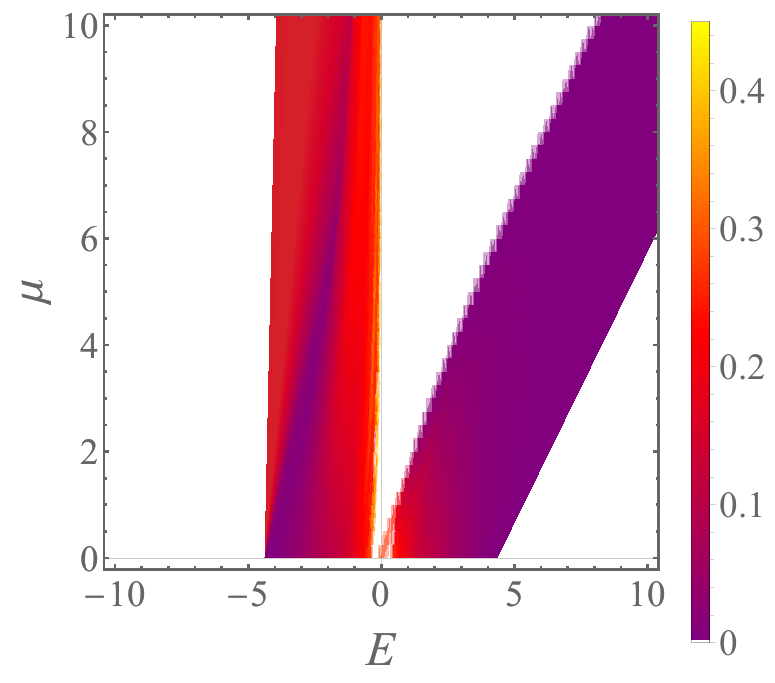}    
    \caption{
    Participation numbers $P_{\mathcal{C}}$ of the \emph{constant shift} potential and their ($E$, $\mu$) dependence for disorder strength (a) $W=1$, (b) $2$ and (c) $5$ as in Fig.\ \ref{fig:probs_l31_CP2}.
    Panels (d-f) are similar to (a-c) but for the $P_{\mathcal{L}}$ of Lieb sites. Color ranges between $P_{\mathcal{C}}$ and $P_{\mathcal{L}}$ are different. As in Fig.\ \ref{fig:probs_l31_CP2}, a white color marks the absence of states.
    }
    \label{fig:Pn_l31_CP0}
\end{figure}

From the ordered checkerboard, let us at last simply flip upwards all the negative potential terms and consider the constant shift case $\varepsilon_\mathcal{C}(\vec{r})= +\mu$ in the cube sub-lattice $\mathcal{C}$ as shown in Fig.~\ref{fig:dir_to_ord}(c). 
This potential keeps the minimal four-site unit cell of Fig.~\ref{fig:Lieb_schematic}(a) intact, while the shift $\mu$ deforms the dispersive bands $E_{D+}$, $E_{D-}$ in Eq.~\eqref{eq:bands} to 
\begin{equation}
    E_{D\pm}
    = \frac{\mu}{2} \pm \frac{1}{2} \sqrt{24 + \mu^2+8(\cos k_x + \cos k_y + \cos k_z)} .
\label{eq:bands_cp0}
\end{equation} 
The resulting spectrum is asymmetric with respect to the macroscopic degeneracy $E = 0$ (cp.\ Fig.\ \ref{fig:probs_l31_CP0}).
On the one hand, the shift $\mu$ opens a gap between the dispersive band $E_{D+}$ and the flat bands $E_{F1,F2}=0$. At the $R = (\pi, \pi, \pi)$ point, where there is band touching for $\mu=0$ (cp.\ Fig.~\ref{fig:Lieb_schematic}(b)), the gap $\Delta E$ grows precisely as $\Delta E \propto \mu$. 
On the other hand, $\mu$ squeezes the dispersive band $E_{D-}$ towards the flat bands $E_{F1}$, $E_{F2}$. 
Hence, with respect to the ordered checkerboard, the potential shift induces only a single band gap, resulting in a loss of the spectral mirror symmetry around $E=0$. 

Regardless of this asymmetry, the shift potential still separates the states in two families as shown in Fig.~\ref{fig:probs_l31_CP0}. 
Indeed, as $\mu$ grows, for each value of the perturbation strength $W$ we observe that 
those states in the positive part of the spectrum sit in the sub-lattice $\mathcal{C}$, while those states in the negative part of the spectrum sit in $\mathcal{L}$. 
We notice, however, that for $W=5$ [panel (c)] at small values of $\mu$ in both families there appear  regions of opposite projected norm, {\it i.e.}, states mostly projected in  $\mathcal{L}$ within the positive energy states, and vice versa.
Furthermore, as this constant shift potential with the additional uncorrelated potential $\delta_{\vec{r}}^{(c)}$ is not self-averaging, we do not observe the accidentally degenerate eigenstates seen previously in Figs.~\ref{fig:probs_l31_CP2} and \ref{fig:probs_l31_CP1}. 

The participation number $P_\mathcal{C}$ shown in Fig.~\ref{fig:Pn_l31_CP0} (a-c), similarly with both former cases, reveal that the high energy states have low participation number indicating localization, 
while the negative energy states have higher values, indicating less localized states. The participation numbers $P_\mathcal{C,L}$ again decrease as $W$ grows. 
This is particularly visible in Fig.~\ref{fig:Pn_l31_CP0}(d-f) for $P_\mathcal{L}$. As the low energy states are mostly localized in the Lieb sub-lattice $\mathcal{L}$, the differences between the participation number values become more evident, emphasizing the difference in nature of the two families of states.  

\section{Perspectives and Conclusions}
\label{sec:conclusions}

Engineering the eigenstates of a system can be achieved by fine-tuning the strengths of the constituent terms of the system's Hamiltonian \cite{Rontgen2019}, for instance, the hopping terms, the interaction, possible external fields, among others.
Flat band lattices are examples of this procedure, where large families of CLS are obtained by fine-tuning the spatial profile of the hopping network in order to ensure destructive interference outside the domain of a CLS ~\cite{Leykam2018}. 
In this work, we (quantum) engineered the eigenstates of the 3D Lieb $\mathcal{L}_3(1)$ model by modulating its onsite potential.  
Following Ref.~\cite{Liu2022}, we focused on onsite potentials which are strictly zero on the Lieb sub-lattice $\mathcal{L}$ -- hence, preserving its macroscopically degenerate set of CLS --  and non-zero in the complementary cube sub-lattice $\mathcal{C}$. 
We considered three types of potential in $\mathcal{C}$, namely a randomized checkerboard, a simple ordered checkerboard and an even simpler constant-site shift. These have been chosen in order to progressively restore order in the Lieb lattice.
We find that each of these potentials turns about half of the dispersive states into low-energy states which mostly live in the clean $\mathcal{L}$ lattice. The remaining half dispersive states get pushed to higher energies and mostly occupy sites in $\mathcal{C}$. 
In other words, an extensive set of eigenstates can be constructed, via the chosen potentials, to resemble a linear superposition of the CLS on the $\mathcal{L}_3(1)$ lattice. 
Furthermore, each chosen potential yields spectral gaps which separate energetically the two families of states, and which grow proportionally with the potential strength. These gaps are robust to uncorrelated disorder perturbations. 

Our results show that this separation between families of states can be achieved with progressively simpler and experimentally achievable potential choices. 
Furthermore, our results are not bounded to the prototypical lattice considered here, but can be extended to other networks, from the generalized Lieb lattices $\mathcal{L}_d(n)$~\cite{Liu2022} to chiral flat band networks~\cite{Ramachandran2017} and beyond. 
The capacity of designing features of extensive sets of eigenstates while tuning their energy finds applications in photonics. For instance, this can be useful in the implementation of hardware realizations  for quantum information storage~\cite{Brehm2021WaveguideQubits}. We speculate that such families of states can be employed in the design of systems of optical lattices for quantum memory and quantum computations~\cite{Arakawa2003Z_NMemory,Nunn2010QuantumLattice}.

\backmatter

\bmhead{Acknowledgments}
Calculations were performed using the Sulis Tier 2 HPC platform hosted by the Scientific Computing Research Technology Platform at the University of Warwick. Sulis is funded by EPSRC Grant EP/T022108/1 and the HPC Midlands+ consortium.
We thank Warwick's Scientific Computing Research Technology Platform for further computing time and support. UK research data statement: Data accompanying this publication are available at 
\href{https://wrap.warwick.ac.uk/178142}{https://wrap.warwick.ac.uk/178142}.




\end{document}